\documentclass[twocolumn,showpacs,aps,prd,10pt,superscriptaddress]{revtex4-1}
\usepackage{graphicx}
\usepackage{dcolumn}
\usepackage{amsmath}
\usepackage{epsfig}
\usepackage{subfigure}
\usepackage{epstopdf}
\usepackage{multirow}
\usepackage{mathrsfs}
\usepackage{colortbl}
\usepackage[bookmarks=false] {hyperref}
\usepackage{cleveref}
\usepackage{lineno}
\usepackage{ulem}
\usepackage{enumerate}
\usepackage{subfigure}
\usepackage{graphicx}

\def\degree{${}^{\circ}$}
\begin{document}


\title{Search for invisible decay of a Higgs boson produced at the CEPC}
\date{\today}

\author{Yuhang Tan}
\affiliation{Institute of High Energy Physics, Beijing 100049, China}
\affiliation{School of Physical Sciences, University of Chinese Academy of Science (UCAS), Beijing 100049, China}
\affiliation{State Key Laboratory of Particle Detection and Electronics, 19B Yuquan Road, Shijingshan District, Beijing 100049, China}
\author{Xin Shi}
\affiliation{Institute of High Energy Physics, Beijing 100049, China}
\affiliation{State Key Laboratory of Particle Detection and Electronics, 19B Yuquan Road, Shijingshan District, Beijing 100049, China}
\author{Ryuta Kiuchi}
\affiliation{Institute of High Energy Physics, Beijing 100049, China}
\affiliation{State Key Laboratory of Particle Detection and Electronics, 19B Yuquan Road, Shijingshan District, Beijing 100049, China}
\author{Manqi Ruan}
\affiliation{Institute of High Energy Physics, Beijing 100049, China}
\affiliation{State Key Laboratory of Particle Detection and Electronics, 19B Yuquan Road, Shijingshan District, Beijing 100049, China}
\author{Maoqiang Jing}
\affiliation{Institute of High Energy Physics, Beijing 100049, China}
\affiliation{School of Physical Sciences, University of Chinese Academy of Science (UCAS), Beijing 100049, China}
\affiliation{State Key Laboratory of Particle Detection and Electronics, 19B Yuquan Road, Shijingshan District, Beijing 100049, China}
\author{Xin Mo}
\affiliation{Institute of High Energy Physics, Beijing 100049, China}
\affiliation{State Key Laboratory of Particle Detection and Electronics, 19B Yuquan Road, Shijingshan District, Beijing 100049, China}
\author{Xinchou Lou}
\affiliation{Institute of High Energy Physics, Beijing 100049, China}
\affiliation{School of Physical Sciences, University of Chinese Academy of Science (UCAS), Beijing 100049, China}
\affiliation{State Key Laboratory of Particle Detection and Electronics, 19B Yuquan Road, Shijingshan District, Beijing 100049, China}
\affiliation{Department of Physics, University of Texas at Dallas, Texas 75080-3021, USA}
\author{Gang Li}
\affiliation{Institute of High Energy Physics, Beijing 100049, China}
\affiliation{State Key Laboratory of Particle Detection and Electronics, 19B Yuquan Road, Shijingshan District, Beijing 100049, China}
\author{Kaili Zhang}
\affiliation{Institute of High Energy Physics, Beijing 100049, China}
\affiliation{School of Physical Sciences, University of Chinese Academy of Science (UCAS), Beijing 100049, China}
\affiliation{State Key Laboratory of Particle Detection and Electronics, 19B Yuquan Road, Shijingshan District, Beijing 100049, China}
\author{Susmita Jyotishmati}
\affiliation{Department of Physics, University of Texas at Dallas, Texas 75080-3021, USA}

\begin{abstract}
The existence of dark matter has been established in astrophysics. However, there is no candidate for DM in the Stand Model (SM). In SM, the Higgs boson can only decay invisibly via $H\rightarrow ZZ^\ast \rightarrow \nu\overline{\nu}\nu\overline{\nu}$ or DM, so any evidence of invisible Higgs decay that exceeds BR (H$\rightarrow$inv.) will immediately point to a phenomenon that is beyond the standard model (BSM). In this paper, we report on the upper limit of BR (H$\rightarrow$invisible) estimated for three channels, including two leptonic channels and one hadronic channel, under the assumption predicted by SM. With the SM ZH production rate, the upper limit of BR (H$\rightarrow$inv.) could reach 0.24\% at the 95\% confidence level.
\end{abstract}

\pacs{XXXXX}

\maketitle

\section{Introduction}
\begin{figure}[ht]
\begin{minipage}[ht]{0.49\linewidth}
\includegraphics[width=1.0\textwidth]{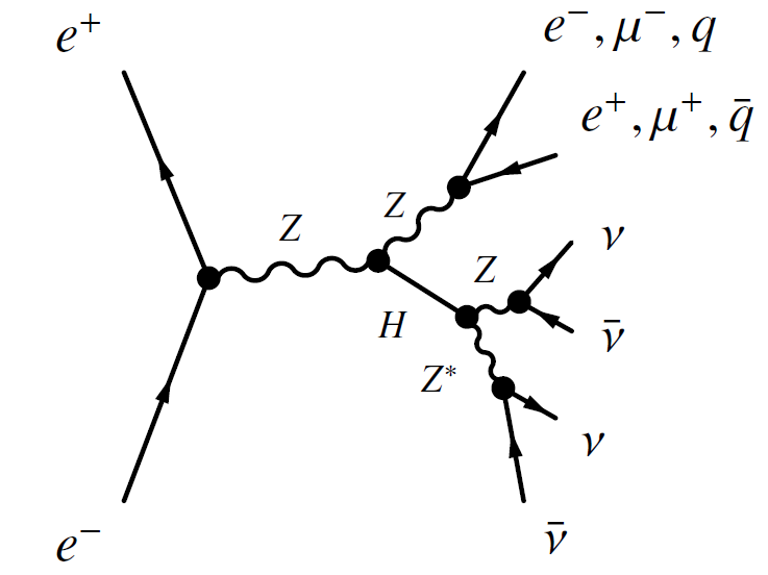}
\end{minipage}
\hfill
    \caption{Feynman diagrams of Higgs boson invisible decay at the CEPC. In the process $e^+e^- \rightarrow$ZH, the invisible decay of Higgs boson will via $H\rightarrow ZZ^\ast \rightarrow \nu\overline{\nu}\nu\overline{\nu}$. Meanwhile, the Z boson decays into leptons or quarks, which are called a lepton or hadron channel, respectively.
    \label{fig:hinvi}}
\end{figure}

Many cosmological evidence have pointed that dark matter (DM) exists in the universe, such as rotation curves in galaxies, galaxy clusters mass evaluation and gravitational lenses in galaxy \cite{bertone2018history,clowe2006direct}. However, there is no candidate for DM in the Stand Model (SM). In collider physics, the Higgs portal model points the only interactions of the DM field are via Higgs field \cite{djouadi2012implications,shrock1982invisible}. The Higgs field might be the portal between the DM sector and the SM sector. One of the methods is to directly search the Higgs decay into DM, where DM is interacting weakly with ordinary matter. Therefore the DM particles produced by the Higgs decay will be completely invisible in the detector. In SM, the Higgs boson can only decay invisibly via $H\rightarrow ZZ^\ast \rightarrow \nu\overline{\nu}\nu\overline{\nu}$, DM, or the fourth generation neutrino~\cite{Belotsky2002}, as shown in Fig.~\ref{fig:hinvi} and its branching ratio is 1.06$\times10^{-3}$. Therefore, any evidence of invisible Higgs decay that exceeds this branching ratio will immediately point to a phenomenon that is beyond the standard model. The invisible decay of the Higgs boson will be a sensitive probe for new physics.

The search for the invisible decay of the Higgs boson has been performed at the Large Hadron Collider (LHC). The signature for the invisible Higgs decay at the LHC is a large missing transverse momentum recoiling against a visible system. The result at the ATLAS and CMS gives an upper limit of 26\% \cite{aaboud2019combination} and 19\%\cite{sirunyan2019search} on the Higgs boson invisible branching ratio BR (H$\rightarrow$inv.). Because the hadron colliders always suffer from a huge number of backgrounds, it is tough to even for the future high luminosity LHC to measure the DM directly. However, it is possible for the precision measurement at an electron-positron collider. The electron-positron Higgs factory is an essential machine for understanding the nature of the Higgs boson. Compared with the LHC, the Higgs production cross-section is available with the recoil mass method without tagging the Higgs decays. In this way, the property of the Higgs boson can be measured precisely without reconstructing the Higgs boson by its decaying products. Therefore, the Higgs boson production can be disentangled in a model independent way. Moreover, the lepton collider has a cleaner environment than the hadron collider which allows better exclusive measurements of Higgs boson decay channels. \par

The Circular Electron Positron Collider (CEPC) \cite{an2019precision} is a Higgs factory proposed by the Chinese high energy physics community. CEPC is designed to deliver a combined integrated luminosity of $5 ab^{-1}$ to two detectors in 7 years. Over one million Higgs boson events will be produced during this period, and it will operate at a center-of-mass energy $\sqrt{s}$ $\sim$ 240-250 GeV. Benefiting from these large statistics, also combined with less energy spread and a more effective particle flow algorithm, the high precision measurements of the mass and width of the Higgs boson can achieve. The upper limit of BR (H$\rightarrow$inv.) in \cite{an2019precision} on CEPC is based on version 1 of CEPC, which the reconstruction algorithm and samples are different from the paper. Therefore, the two results are less comparable.\par

The paper performs three independent analyses corresponding to $\mu\mu$H, $ee$H and qqH channels to estimate the upper limit of BR (H$\rightarrow$inv.) measurement at the CEPC. This paper is organized as follows: Section II is a brief introduction of the CEPC detector and Monte Carlo simulation. Section III is the introduction of the event selection of three channels. How to get the result of the upper limit and the dependency of the Boson Mass Resolution (BMR) will be discussed in section IV. Section V is the conclusion.
\section{Detector and Monte Carlo simulation}
The CEPC detector is designed by using the Inter-national Large Detector (ILD) \cite{abe3396international,behnke2013international} as a reference, and a physics program of the CEPC is the precision measurements of the Higgs boson properties. The CEPC detector should reconstruct and identify all key physics objects including charged leptons, photons, jets, missing energy and missing momentum. To reach this goal, the detector of CEPC is simulated using M\textsc{okka} \cite{mora2002detector} and Geant4 \cite{sartini2010nuclear}. The simulation contains full simulation and slimmed reconstruction samples without hits. The CEPC-v4 detector is comprised of the tracking system, a Time-Projection-Chamber tracker (TPC), a high granularity calorimeter system, a solenoid of 3 Tesla magnetic field, and a muon detector embedded in a magnetic field return yoke. The tracking system consists of silicon vertexing and tracking detectors. The calorimetry system consists of electromagnetic calorimeter (ECAL) and Iron-Scintillator for the hadronic calorimeter (HCAL)~\cite{an2019precision}. \par
The analysis is performed on Monte Carlo (MC) samples simulated on the CEPC conceptual detector. The Higgs boson signal and SM background at a center-of-mass energy of 240 GeV, corresponding to an integrated luminosity of 5.6 $ab^{-1}$, are generated with W\textsc{hizard}1.95~\cite{kilian2011whizard}. The generated events are then processed with M\textsc{okkac} \cite{mora2002detector}, and attempt to reconstruct all visible particles with A\textsc{rbor} \cite{Ruan:2014paa}. All samples are grouped into signal and background and then classified according to their final states. The Higgs boson signal and part of the leading backgrounds are processed with Geant4 based full detector simulation and reconstruction. The rest of the backgrounds are simulated with a dedicated fast simulation tool, where the detector acceptance, efficiency and intrinsic resolution for different physics objects are parameterized.\par
The cross-sections of major Standard Model processes of $e^{+}e^{-}$ collisions as a function of center-of-mass energy $\sqrt{s}$ were used in the simulation, including Higgs production as well as the major backgrounds, where the ISR effect has been taken into account. For the signal, the article mainly focuses on the process of $e^{+}e^{-}\rightarrow$ ZH, which will be called a “ZH” process. Then the Higgs will decay into four neutrinos, and Z bosons will decay into leptons or hadrons. This analysis only chooses three channels: Z$\rightarrow$ee, Z$\rightarrow$qq and Z$\rightarrow \mu\mu$ as signal channels. For the background, the major SM background is divided into the 2-fermion processes and the 4-fermion processes according to the final status. The 2-fermion backgrounds are $e^{+}e^{-}$$\rightarrow $$f\overline{f}$ where f refers to all lepton and quark pairs except $t\overline{t}$. The 4-fermion backgrounds are divided as ZZ, WW, ZZorWW, Single Z, Single W, Single Z or Single W. The four fermion in the final states can be combined into two intermediate bosons. If they are two Z bosons or two W bosons, the processes can be named as “ZZ” or “WW” accordingly. For the process whose final states contains a pair of electron and the accompanying electron neutrino, they will be excluded from the ZZ and WW group and named as “single Z” or “single W”. The “single Z” include $(e^-,e^+)$ or $(\nu_{e},\nu_{e})$, and the “single W" include $(e,\nu_{e})$. Some final particles can come from “ZZ” and “WW”, and those processes will be named as “ZZ or WW”. The “single Z or single W” is similar to “ZZ or WW”, and the final particles include $e^-$, $e^+$, $\nu_{e}$, $\nu_{e}$ which can come from “Single Z” and “Single W”.\par
\section{event selection}
As mentioned in section I, to improve the precision of Higgs, this article uses the recoil mass method, which doesn't use the information of the Higgs boson decay, and the method is model independent. The signal of this analysis includes three different channels, including ZH (Z$\rightarrow$ee, H$\rightarrow$inv.), ZH (Z$\rightarrow$qq, H$\rightarrow$inv.) and ZH (Z$\rightarrow \mu\mu$, H$\rightarrow$inv.). Table.~\ref{tab:higgssignal} shows detailed information of the signal channels. The muon and electron can easily be identified, and their momentum can be precisely measured in the detector. By tagging the muon or electron or total visible particles from the associated Z boson decays, the signal events can be reconstructed with the recoil mass method. The signal events form a peak in the $M_{recoil}^{Z}$ distribution at the Higgs boson mass, which is about 125 GeV.
In event selection, to better select the signal samples, this analysis assumes the BR (H$\rightarrow $inv.)=50\%. Below is the detailed event selection for each channel.
 \begin{table}[hbtp]
 \caption{Cross sections of the Higgs boson production at $\sqrt{s}$=240 GeV  and numbers of events expected in 5.6 $ab^{-1}$.}
 \label{tab:higgssignal}
 \small
 \begin{center}
 \renewcommand{\arraystretch}{1.2}
 \begin{tabular}{ccc}
 \hline
process & cross sections (fb) & expected\\ \hline
 ffH &203.66& 1140496 \\
 $e^{+}e^{-}$H &7.04 & 39424\\
${\mu^+}{\mu^-}$H &6.77&37912\\
$\tau^{+}\tau^{-}$H &6.75& 37800 \\
$\nu\overline{\nu}$H &46.29 &259224\\
$q\bar{q}$H &136.81&766136 \\
 \hline
 \end{tabular}
 \end{center}
 \end{table}
\subsection{ZH (Z$\rightarrow$qq,H$\rightarrow$inv.)}
In ZH (Z$\rightarrow$qq, H$\rightarrow$inv.) process, due to the presence of quarks, there will be many final states. The event selection can use the information of visible particles or jets. The jet is a narrow cone of hadrons and other particles produced by the hadronization of a quark or gluon. The principle of the event selection is based on the distribution of the signal and background at different selection parameters which can give the direct range of this selection parameter, and the signal strength accuracy: $\frac{\sqrt{S+B}}{S}$ which can accurately judge the selection parameter range further from its value. The comprehensive event selections are following: The recoil mass of all visible particles is the Higgs mass. Considering the resolution of the detector, the $M_{recoil}^{visible}$ is limited to (120,150) GeV. To suppress 2-fermion backgrounds, the transverse momentum of visible particles is required to satisfy 18 GeV$<P_{T}^{visible}<$60 GeV and the difference of the azimuth angles of the two jets should be less than 175\degree. The visible energy is the energy of two quarks which can be described as: 
\begin{equation}
\label{E3}
   (M_{rec}^{visible})^{2} = (\sqrt{s}-E_{visible})^{2}- P_{visible}^{2} 
\end{equation}
\begin{equation}
\label{E4}
M_{visible}^{2} = E_{visible}^{2} - P_{visible}^{2}
\end{equation}. \par
\begin{figure*}[t]
\begin{minipage}[t]{0.48\linewidth}
\includegraphics[width=1.0\textwidth]{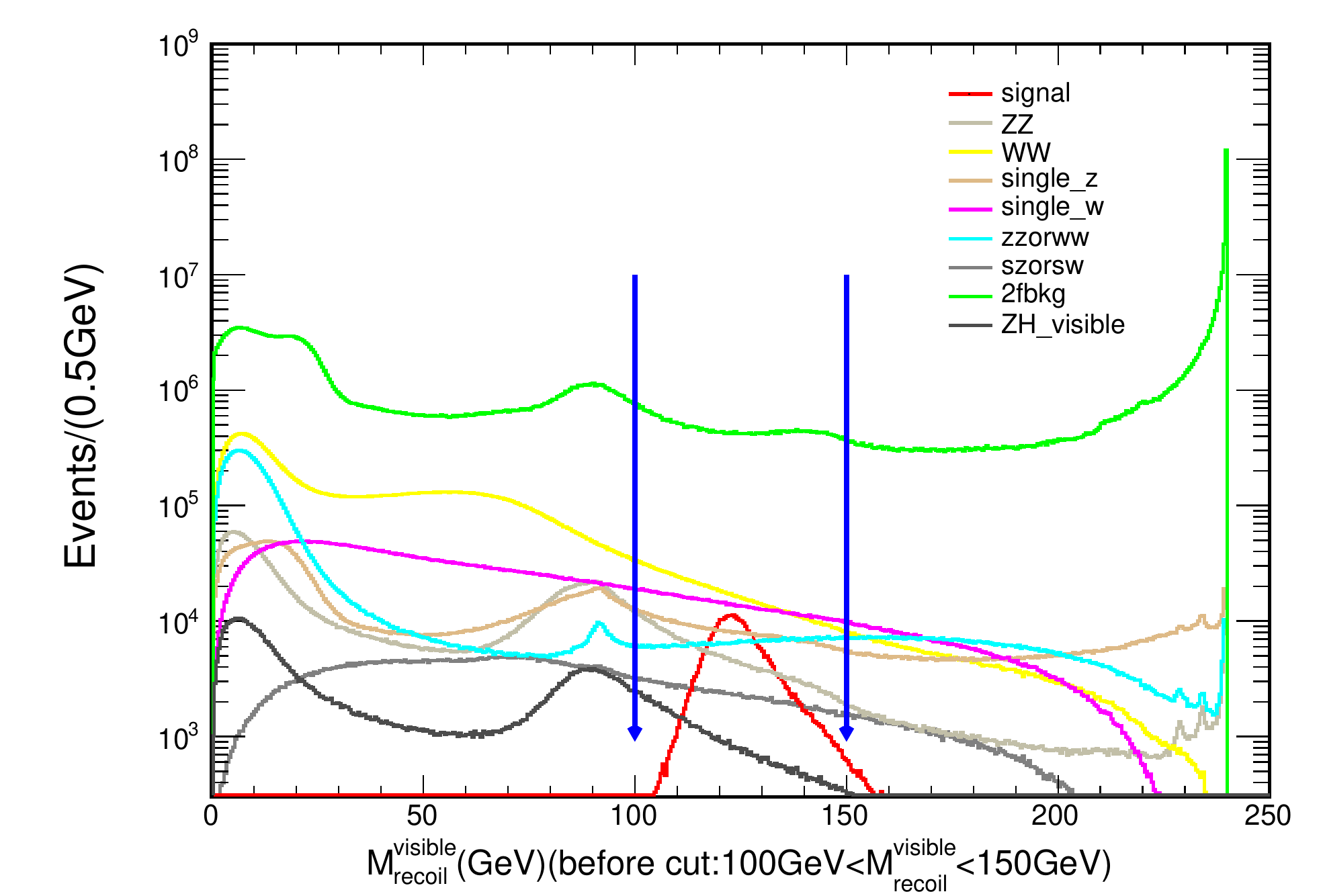}
\end{minipage}
\hfill
\begin{minipage}[t]{0.48\linewidth}
\includegraphics[width=1.0\textwidth]{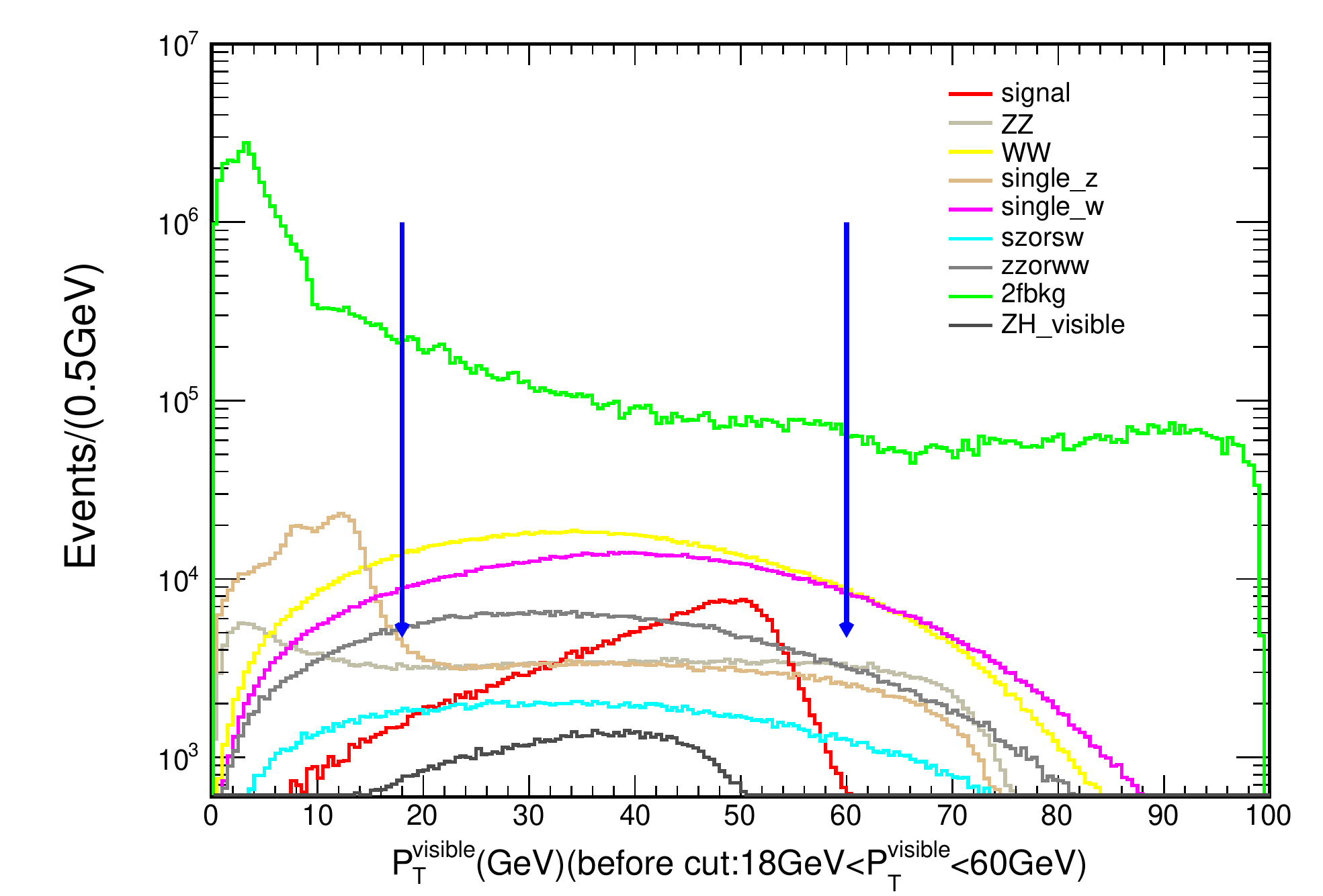}
\end{minipage}
\hfill
\begin{minipage}[t]{0.48\linewidth}
\includegraphics[width=1.0\textwidth]{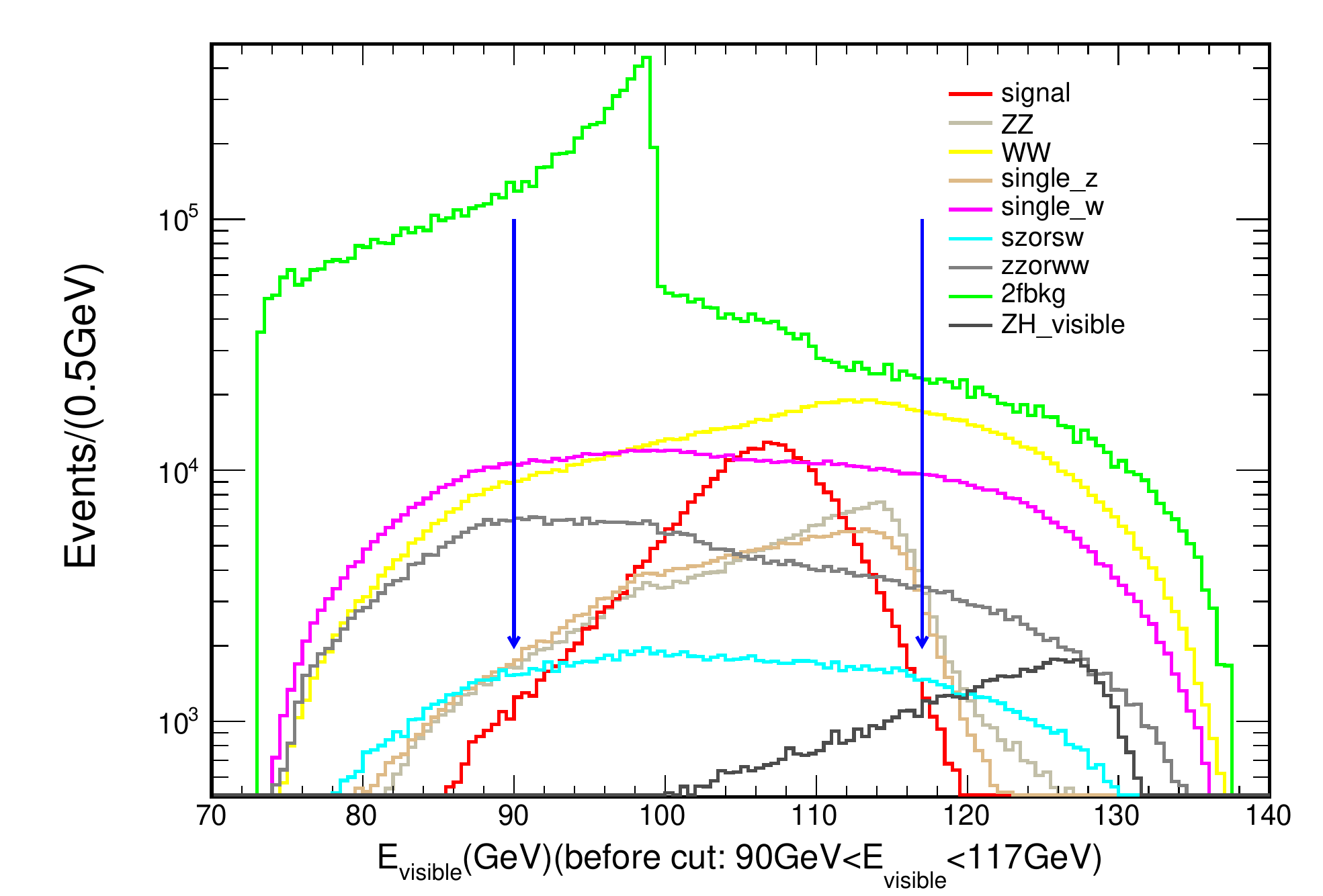}
\end{minipage}
\hfill
\begin{minipage}[t]{0.48\linewidth}
\includegraphics[width=1.0\textwidth]{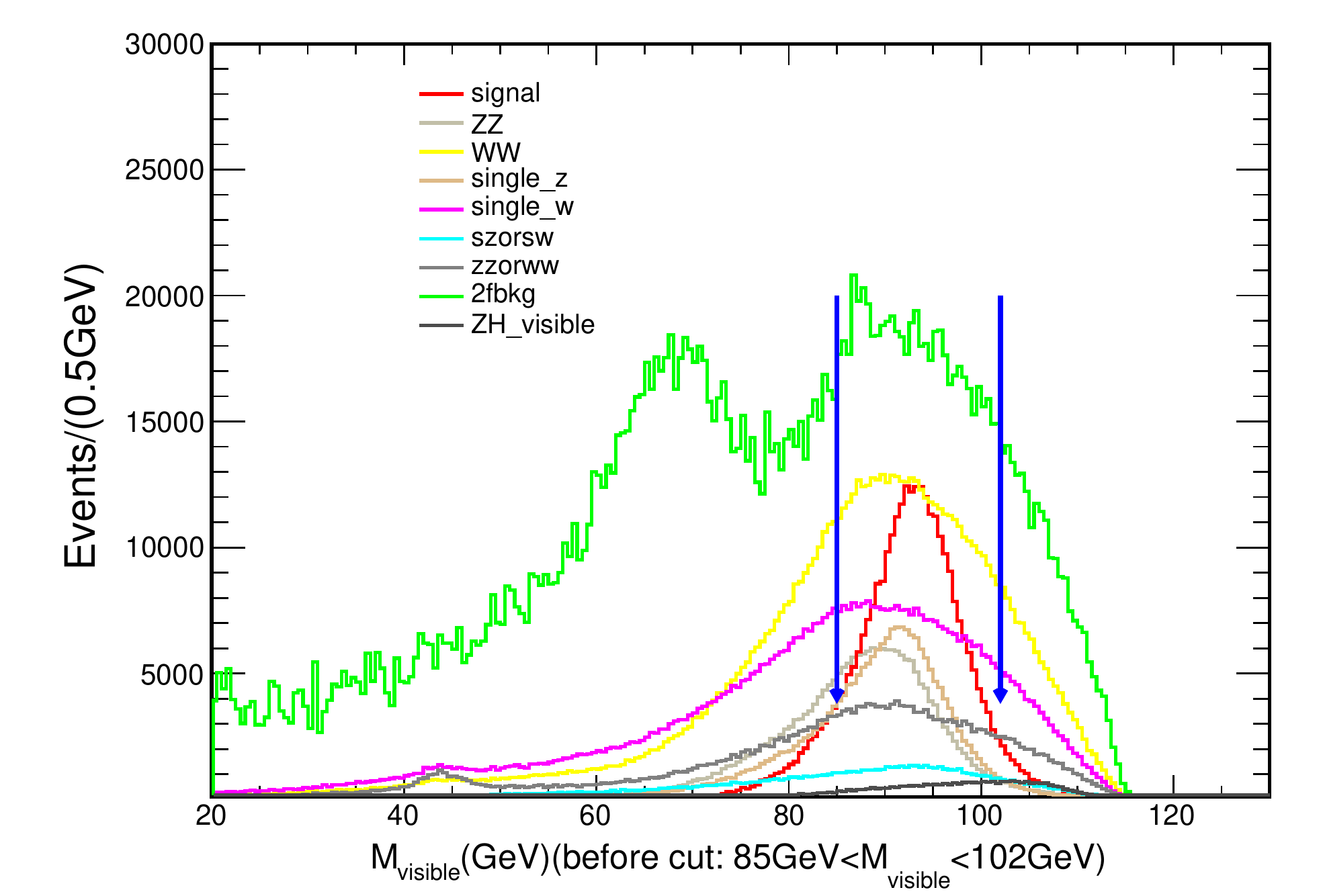}
\end{minipage}
\hfill
\caption{(color online) The distribution of $M_{recoil}^{visible}$, $P_{T}^{visible}$, $E_{visible}$ and $M_{visible}$ for signal and backgrounds before the cut its own (Based on Table.~\ref{tab:all channel are grouped  cutq}). The blue arrows are cut range.
\label{fig:qqH}} 
\end{figure*}

In the equation (\ref{E3}) and equation (\ref{E4}), the $M_{Higgs}=125$ GeV, $\sqrt{s}=240$ GeV, and the $M_{visible}$ is approximately equal to the mass of Z boson. From the equations of the recoil mass and invariant mass, the $E_{visible}$ should be near 105 GeV, and $P_{visible}$ should be near 52 GeV. The invariant mass $M_{visible}$ closer to the Z boson mass which is 91.2 GeV, and $M_{visible}$ is limited to (85,102) GeV. Due to the presence of quarks, the final state particles may include many neutral particles. So it is necessary to limit the minimum number of neutral particles. The number of neutral less than 15 is selected by comparing the value of the signal strength accuracy.
Meanwhile, the leptons isolated to the jets of quarks are selected to suppress the background containing two quarks and leptons. These leptons are called isolated particles, and the number of isolated muons and isolated electrons are zero in the signal channel. Fig.~\ref{fig:qqH} shows the distribution of the signal and background on different cut parameters, which can roughly determine the range of each cut parameter. \par
Table.~\ref{tab:all channel are grouped cutq} shows the yields of signal (qqH) and backgrounds of the cut chain. Since the BR (H$\rightarrow$inv.)is assumed as 50\%, and the cross-section of ZH (Z$\rightarrow$qq, H$\rightarrow$inv.) is 136.81 fb, the number of signal samples is 383068. Moreover, the value of the signal strength accuracy $\frac{\sqrt{S+B}}{S}$ is used to judge the efficiency of the cut in each step, and the table also shows the number of remaining backgrounds and signals. After the event selection, the signal selection efficiency is 60.40\%, and the total background rejection efficiency is 99.98\%. The main backgrounds left are the channels containing four particles: two neutrinos and two quarks accounted for 50\%, and the channels containing four particles: tau, neutrino, u quark and d quark accounted for 31\%. Moreover, the remaining backgrounds are similar to the signal channel, it is hard to suppress further.
 \begin{table*}[hbtp]
 \caption{Yields for backgrounds and ZH (Z$\rightarrow$qq,H$\rightarrow$inv.) signal at the CEPC, with $\sqrt{s}$=240 GeV and integrated
luminosity of 5.6 $ab^{-1}$. (Assume BR (H$\rightarrow$inv.)=50\%)}
 \label{tab:all channel are grouped  cutq}
  \tiny
 \begin{center}
 \renewcommand{\arraystretch}{1.2}
 \begin{tabular}{cccccccccccccc}
 \hline
 Process&  qqh\_inv. &2f &single\_w &single\_z &szorsw &zz&ww &zzorww&ZH\_visible &total\_bkg&$\frac{\sqrt{S+B}}{S}$ \\ \hline
Total generated & 383068 &801152072 &19517400  &9072952&1397088 &6389429 &50826213&20440840 &1140496&909936490&7.88 \%   \\
100GeV$<M_{recoil}^{visible}<$150GeV & 369001&47294921 &1388874&822725 &229216 &507558 &1752824 &658200  &97384&52751702& 1.98 \%   \\
18GeV$<P_{T}^{visible}<$60GeV & 335572  &9165308&1000761 &269323 &152273&282624 &1294263 &462027 &79965&12706544& 1.08 \%    \\
90GeV$<$$E_{visible}$$<$117GeV & 319558 &5748711 &595694&223044  &92958 &231050&785389 &272515 &33705 &7983066& 0.90 \%   \\
85GeV$<M_{visible}<$102GeV & 268930 &605788 &238190&148842 &39280  &135635 &392275&113043 &18282&1691335& 0.52 \%    \\
$\Delta\phi_{dijet}<175$\degree& 259553 &390075&230271 &141490 &38358&129130  &379928 &109734 &17393&1436379& 0.50 \%    \\
30GeV$<P_{visible}<$58GeV& 242860 &241508&148607  &69450 &24392&46800 &226881 &74780 &13465 &845883& 0.43 \%   \\
$N_{neutral}>15$& 242341 &18081&22594 &64324  &149&44338 &128425 &8616 &11852&298379& 0.30 \%     \\
$N_{IsoMuon}=0,N_{IsoElectron}=0$& 231374 &8423&9604 &60645 &28 &41536 &76617 &6447 &9219&212519& 0.29 \%  \\
 Efficiency& 60.40 \%& 0.00 \% &0.05 \%&0.67 \% &0.00 \%&0.65 \% &0.15 \%&0.03 \% &0.81 \% &0.02 \%&\\
\hline
 \end{tabular}
 \end{center}
 \end{table*}
\subsection{ZH (Z$\rightarrow \mu^+\mu^-$, H$\rightarrow$inv.) and ZH (Z$\rightarrow e^+e^-$, H$\rightarrow$inv.)}
Because the ZH (Z$\rightarrow \mu^+\mu^-$, H$\rightarrow$inv.) process and ZH (Z$\rightarrow e^+e^-$, H$\rightarrow$inv.) process are similar, the two processes will be introduced together. Firstly, it is natural that only a pair of oppositely charged muons or electrons is required in the visible final states. By tagging two muons or two electrons, many related parameters can be used to suppress the backgrounds. The traditional event selections are as follows: The recoil mass of two muons or two electrons is near the Higgs boson mass, and consider the resolution of muon and electron. The recoil mass should satisfy 120 GeV$<M_{recoil}^{\mu^{+}\mu^{-}}<$150 GeV or 120 GeV$<M_{recoil}^{e^{+}e^{-}}<$170 GeV, and the invariant mass of two muons or two electrons closer to Z boson mass. To suppress the 2-fermion backgrounds, the transverse momentum of the muon pair is required to be more than 12 GeV, and the transverse momentum of the electron pair is required to satisfy 12 GeV$<P_{T}<$55 GeV. The $\Delta\phi_{\mu^+\mu^-}$ is less than 175\degree or $\Delta\phi_{e^+e^-}$ is less than 176\degree is required to suppress the 2-fermion backgrounds further. The visible energy is mainly the energy of two muons or two electrons, and the energy of two muons or two electrons can be described as: 
\begin{equation}
\label{E1}
    (M_{rec}^{\mu^+\mu^-/e^{+}e^{-}})^{2} = ( \sqrt{s} - E_{\mu^+\mu^-/e^{+}e^{-}})^{2} - P_{\mu^+\mu^-/e^{+}e^{-}}^{2}  
\end{equation}
\begin{equation}
\label{E2}
M_{Z}^{2} = E_{\mu^+\mu^-/e^{+}e^{-}}^{2} - P_{\mu^+\mu^-/e^{+}e^{-}}^{2}
\end{equation} \par
\begin{figure*}[t]
\begin{minipage}[t]{0.49\linewidth}
\includegraphics[width=1.0\textwidth]{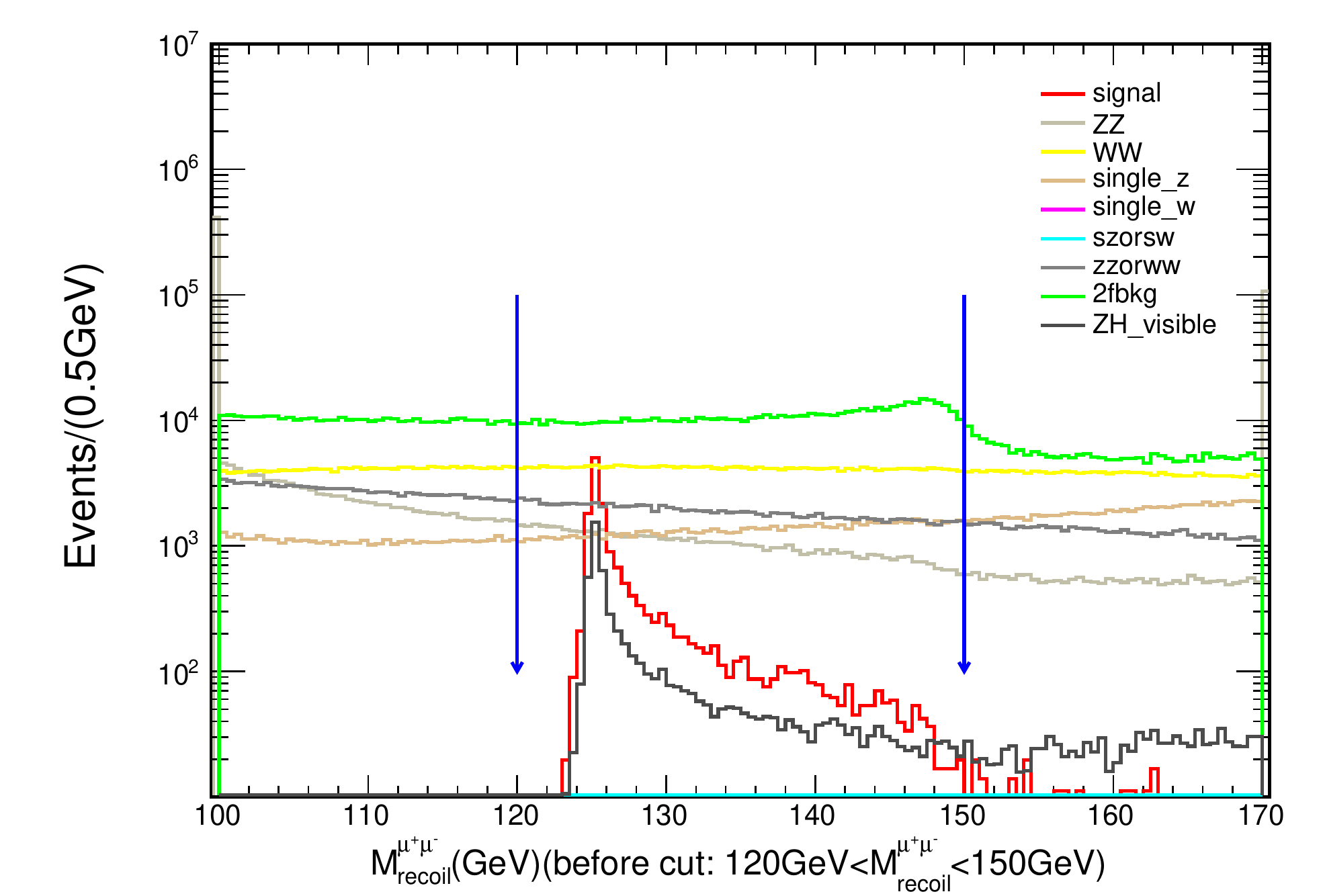}
\end{minipage}
\hfill
\begin{minipage}[t]{0.49\linewidth}
\includegraphics[width=1.0\textwidth]{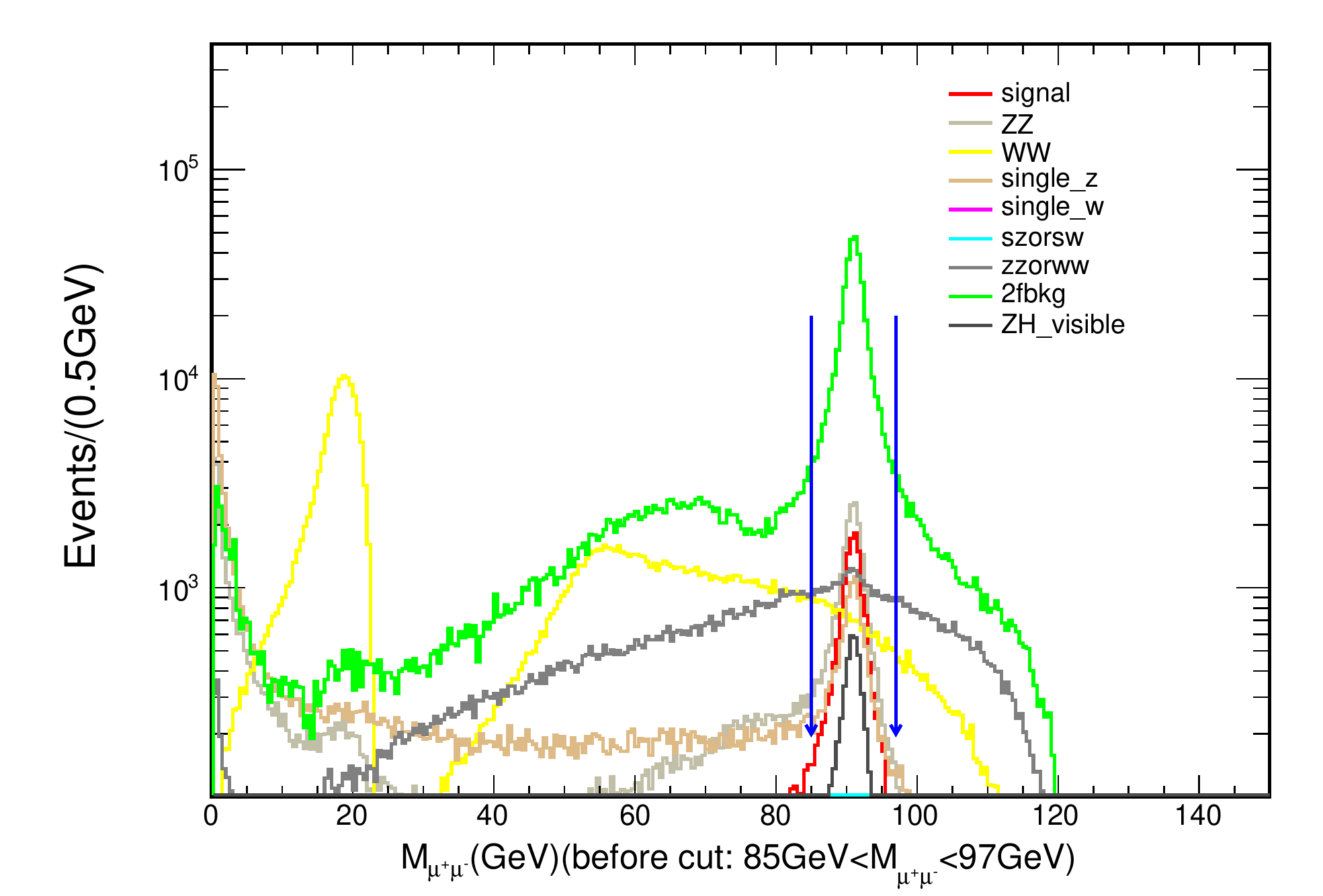}
\end{minipage}
\hfill
\begin{minipage}[t]{0.49\linewidth}
\includegraphics[width=1.0\textwidth]{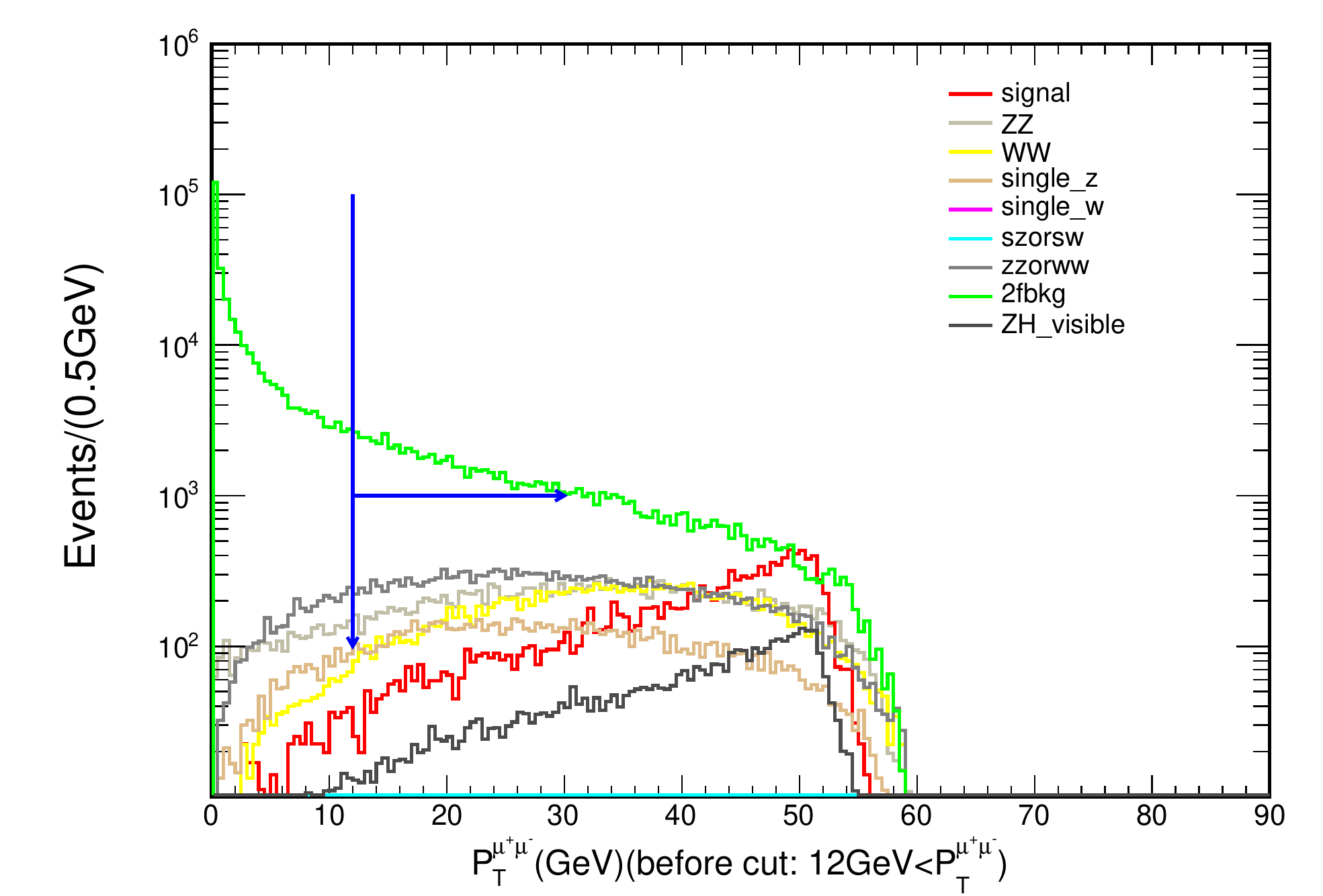}
\end{minipage}
\hfill
\begin{minipage}[t]{0.49\linewidth}
\includegraphics[width=1.0\textwidth]{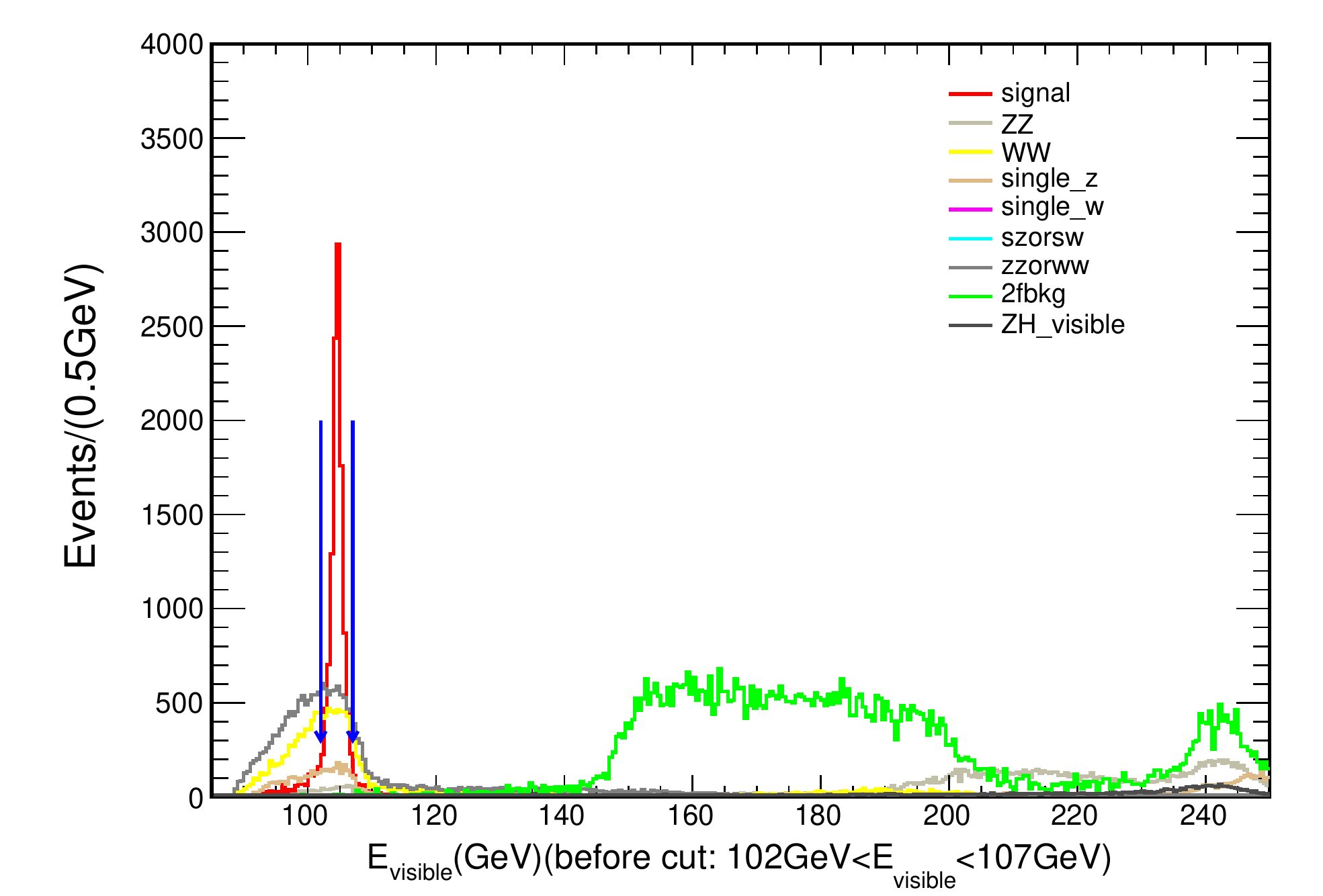}
\end{minipage}
\hfill

\caption{(color online) The distribution of $M_{recoil}^{\mu^+\mu^-}$, $P_{T}^{\mu^+\mu^-}$, $E_{visible}^{\mu^+\mu^-}$ and $M_{\mu^+\mu^-}$ for signal and backgrounds before the cut its own (Based on Table.~\ref{tab:all channel are grouped  cut}). The blue arrows are cut range.
\label{fig:mumuH}}
\end{figure*}
\begin{figure*}[t]
\begin{minipage}[t]{0.49\linewidth}
\includegraphics[width=1.0\textwidth]{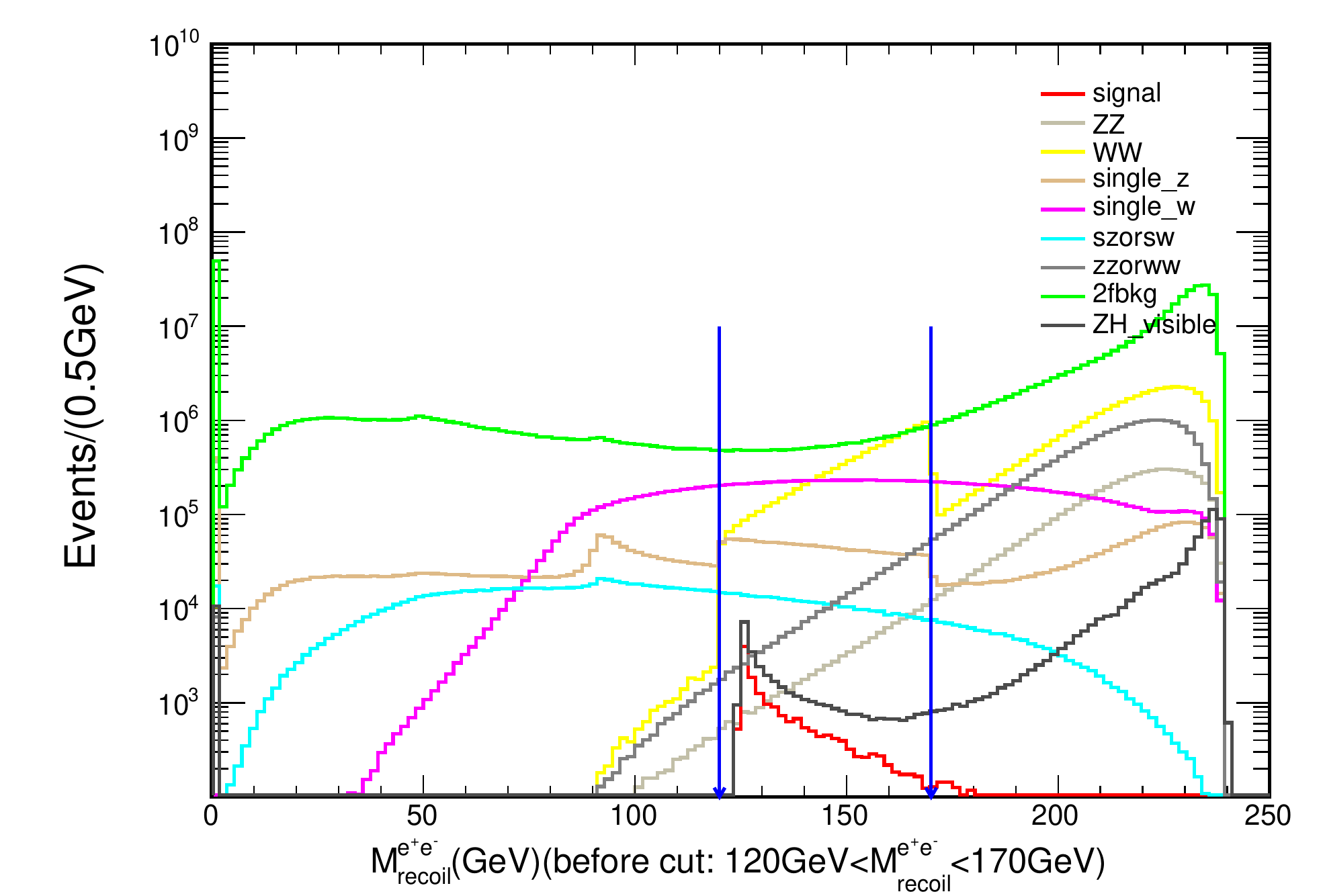}
\end{minipage}
\hfill
\begin{minipage}[t]{0.49\linewidth}
\includegraphics[width=1.0\textwidth]{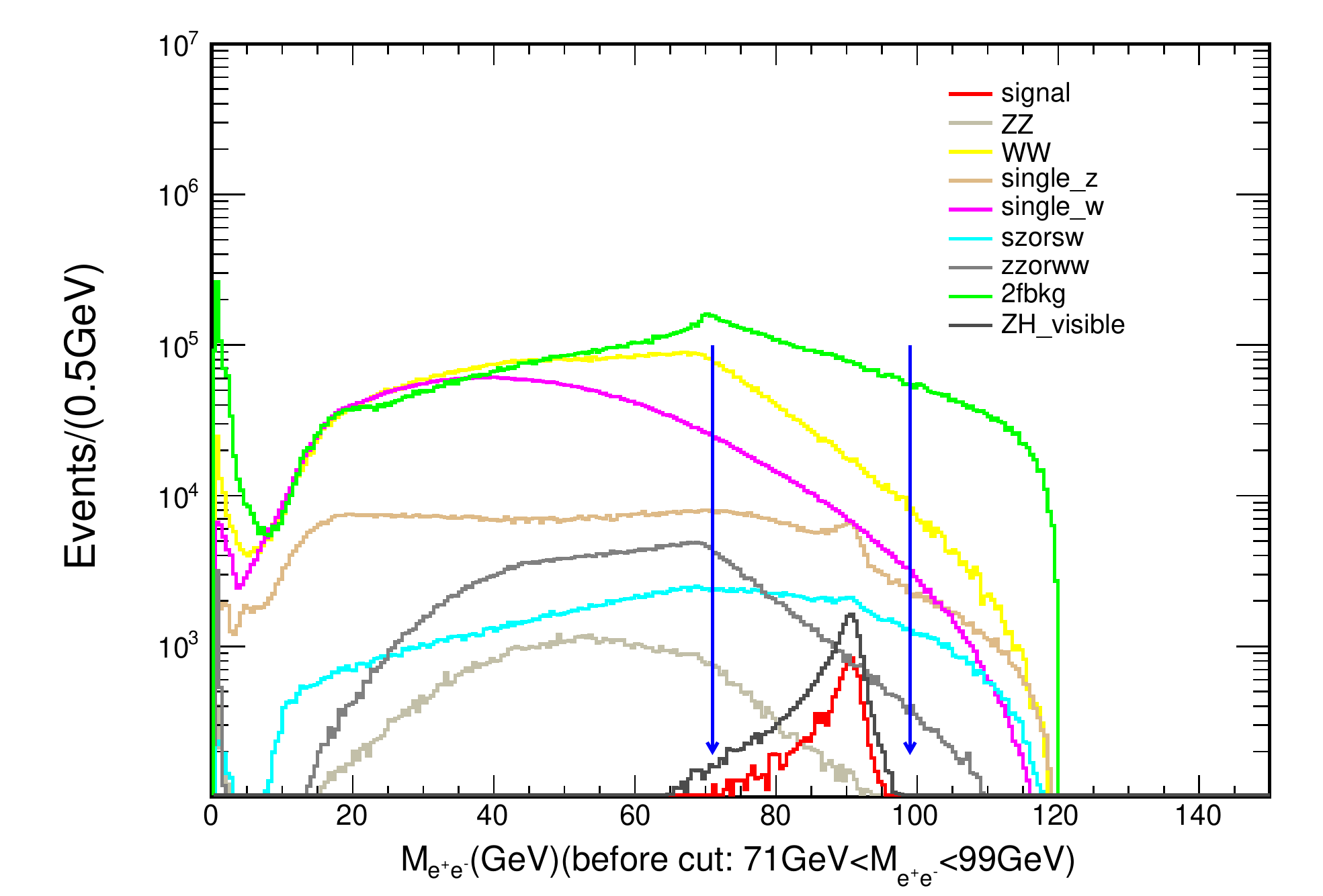}
\end{minipage}
\hfill
\begin{minipage}[t]{0.49\linewidth}
\includegraphics[width=1.0\textwidth]{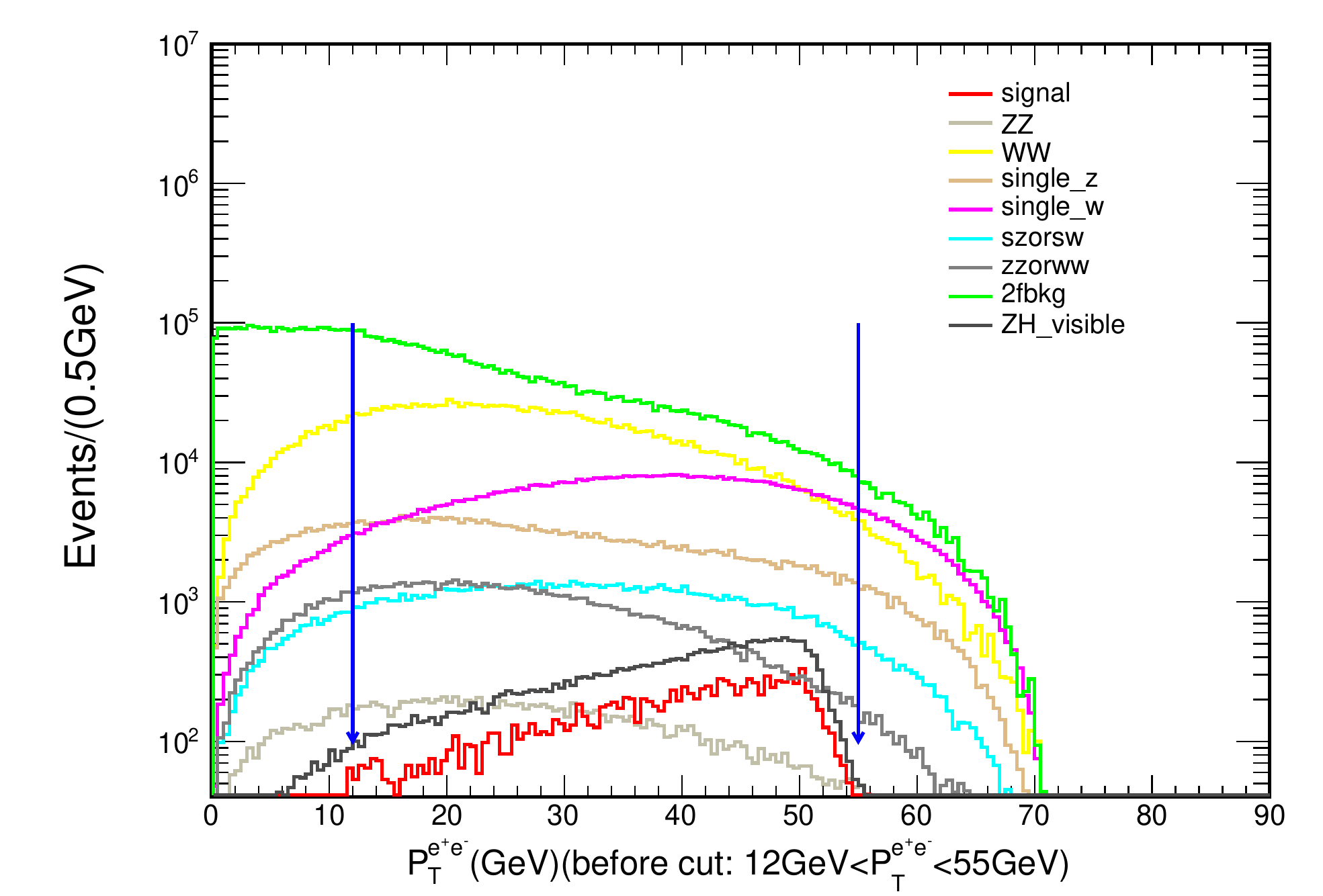}
\end{minipage}
\hfill
\begin{minipage}[t]{0.49\linewidth}
\includegraphics[width=1.0\textwidth]{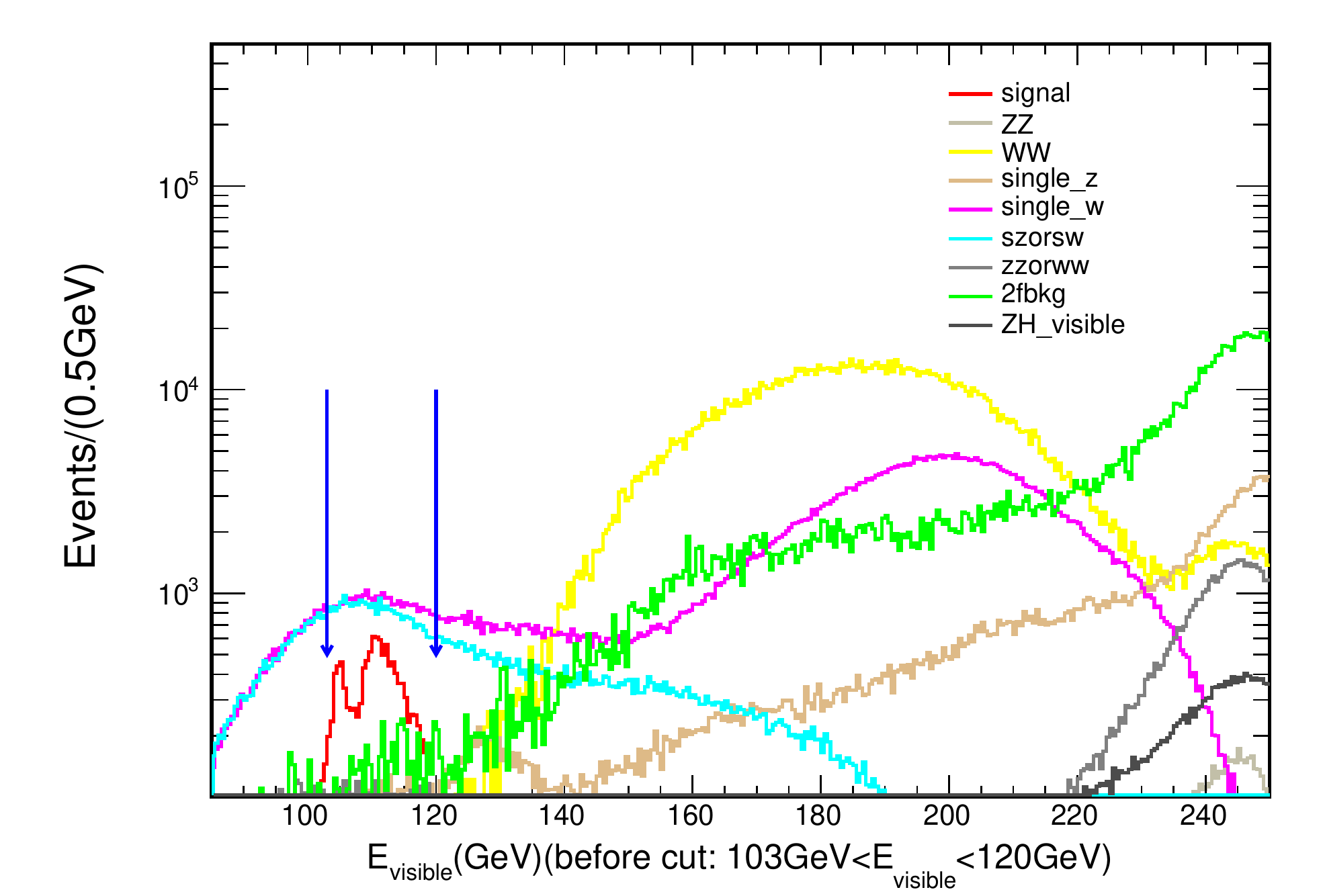}
\end{minipage}
\hfill

\caption{(color online) The distribution of $M_{recoil}^{e^+e^-}$, $P_{T}^{e^+e^-}$, $E_{visible}^{e^+e^-}$ and $M_{e^+e^-}$ for signal and backgrounds before the cut itself (Based on Table.~\ref{tab:all channel are grouped  cute}). The blue arrows are cut range.
\label{fig:eeH}}
\end{figure*}

In the above formula, $E_{\mu^+\mu^-/e^{+}e^{-}}$ is $E_{\mu^+\mu^-}$ or $E_{e^{+}e^{-}}$, and other parameters are the same.
 From equation (\ref{E2}), the value of $E_{visible}^{\mu^+\mu^-}$ or $E_{visible}^{e^+e^-}$ is approximately 105 GeV. From equation (\ref{E1}) and equation (\ref{E2}) calculations, $\frac{E_{\mu^{+}\mu^{-}}}{P_{\mu^{+}\mu^{-}}}$ is less than 2.4, and $\frac{E_{e^+e^-}}{P_{e^+e^-}}$ is limited to (1.8,2.4).
The distribution of $M_{recoil}^{\mu^+\mu^-/e^{+}e^{-}}$, $P_{T}^{\mu^+\mu^-/e^{+}e^{-}}$, $E_{visible}^{\mu^+\mu^-/e^{+}e^{-}}$ and $M_{\mu^+\mu^-/e^{+}e^{-}}$ for signal and backgrounds before the cut its own are shown in Fig.~\ref{fig:mumuH} and Fig.~\ref{fig:eeH}. The range and effect of the cut is evident in these diagrams.\par
In Fig.~\ref{fig:eeH}, there are two peaks at the signal distribution of $E_{visible}^{e^{+}e^{-}}$. Due to the electron and photon will reconstruct into one cluster when the photon generated by the electron is near the electron. The corresponding track of the cluster is identified as the electron, and the energy of the track is the energy of the electron after radiating the photon. In this case, the protection mechanism, which the energy of cluster energy subtract track energy will become a neutral particle, will appear in the reconstruction algorithm. Due to the large energy of the cluster, the energy of these neutral particles will be zero or large value, which corresponds to the neutral particle is identified and unidentified, and this will lead to two peaks at the distribution of $E_{visible}^{e^{+}e^{-}}$. In the channel of ZH (Z$\rightarrow \mu^+\mu^-$, H$\rightarrow$inv.), most muon will not radiate and generate photons, so there are not two peaks at the energy visible distribution in Fig.~\ref{fig:mumuH}. \par

In addition to the common selection conditions, in order to suppress the background containing tau particles and quarks, using the tau information as the cut is studied. The introduction of the specific candidate tau can be found in ~\cite{yutan_2}. The candidate tau, which contains tau, electron and muon, is selected by a series of selection conditions. In ZH (Z$\rightarrow \mu^+\mu^-$, H$\rightarrow$inv.), due to the visible particles of the signal channel is equal to candidate tau, the value of the energy of visible particles minus the energy of candidate tau is around 0 GeV. And in the background channel which contain quarks, this value will be more than 0 GeV. The special parameter is the recoil mass of the visible particles removal the candidate tau, so the parameter $ReM_{visdtau}$ requires larger than 230 GeV.\par
Table.~\ref{tab:all channel are grouped cut} is the yields for backgrounds and ZH (Z$\rightarrow \mu^+\mu^-$, H$\rightarrow$inv.) signal. The cross-section of the signal is 6.77 fb. Assuming BR (H$\rightarrow$inv.) is 50\%, the expected event of signal is 18956. After all cut, $\frac{\sqrt{S+B}}{S}$ is 1.03 \%. The cut of $ReM_{visdtau}>$230 GeV causes $\frac{\sqrt{S+B}}{S}$ to change from 1.28\% to 1.03 \%, which prove that the effect of this selection condition is significant. All the main backgrounds left contain two muons and two neutrinos, which is the same with the signal. Therefore, it is hard to further suppress the background. \par
 \begin{table*}[hbtp]
 \caption{Yields for backgrounds and ZH (Z$\rightarrow \mu^+\mu^-$, H$\rightarrow$inv.) signal at the CEPC, with $\sqrt{s}$=240 GeV and integrated
luminosity of 5.6 $ab^{-1}$. (Assume BR (H$\rightarrow$inv.)=50\%)}
 \label{tab:all channel are grouped  cut}
 \tiny
 \begin{center}
 \renewcommand{\arraystretch}{1.2}
 \begin{tabular}{ccccccccccccc}
 \hline
 Process& $\mu^{+}\mu^{-}$H\_inv. &2f &single\_w &single\_z &szorsw &zz &ww &zzorww &ZH&total\_bkg&$\frac{\sqrt{S+B}}{S}$ \\ \hline
Total generated & 18956 &801152072 &19517400 &9072951 &1397088 &6389430&50826214 &20440840 &1140495&909936490&159.13 \%   \\
$N_{\mu^{+}}=1,N_{\mu^{-}}=1$ & 16851 &22737312 &36122 &723397 &0&702041 &1255610 &1223595&59978&26738055& 30.70 \%   \\
120GeV$<M_{Recoil}^{\mu^{+}\mu^{-}}<$150GeV & 16431 &652616 &24 &81165 &0 &62389 &250796&112141 &5680&1164811& 6.61 \%    \\
85GeV$<M_{\mu^{+}\mu^{-}}<$97GeV & 13957 &381054 &0 &10576&0 &20850 &16718 &24417 &4485 &458100& 4.92 \%   \\
12GeV$<P_{t}^{\mu^{+}\mu^{-}}$ & 13522 &92197 &0 &9333&0&18253  &15903 &21061 &4324&161071& 3.09 \%    \\
$\Delta\phi^{\mu^{+}\mu^{-}}<175$\degree & 12990 &72196 &0 &8754 &0 &17023 &14768&20230 &4136&137107& 2.98 \%    \\
102GeV$<E_{visible}<$107GeV & 11365 &61 &0 &1455&0 &483 &4378 &5434 &9 &11820& 1.34 \%   \\
 $\frac{E_{\mu^{+}\mu^{-}}}{P_{\mu^{+}\mu^{-}}}<2.4$& 11216 &26 &0 &1343  &0&439 &3502&4088 &5&9403& 1.28 \%     \\
 $ReM_{visdtau}>$230GeV& 11143 &26 &0 &1338 &0 &436 &66&52 &4  &1922 & 1.03 \% \\
 Efficiency& 58.78 \%& 0.00 \% &0.00 \%&0.01 \% &0.00 \%&0.01 \% &0.00 \%&0.00 \% &0.00 \% &0.00 \%&\\
 \hline
 \end{tabular}
 \end{center}
 \end{table*}
In the ZH (Z$\rightarrow e^+e^-$, H$\rightarrow$inv.) process, the event selection uses the candidate tau information to suppress the background containing tau particles and quarks. The reason is similar to ZH (Z$\rightarrow \mu^+\mu^-$, H$\rightarrow$inv.) process, and the $ReM_{visdtau}$ large than 220 GeV is required. Apart from this, in order to further suppress the background which only contains the tau particles, introduce a special parameter the position of secondary vertexing. Tau particle will decay into other particles, so the position of its secondary vertexing will be greater than the electron and muon, which can be used to separate electron, muon and tau. The parameter of secondary vertex named impact tau is less than 0.0011 in this paper. \par
 Table~\ref{tab:all channel are grouped cute} is yields for backgrounds and ZH (Z$\rightarrow e^+e^-$, H$\rightarrow$inv.) signal at the CEPC.
 After the cut of $ReM_{visdtau}>$220 GeV and Impact\_{Tau}$<$0.0011, the value of $\frac{\sqrt{S+B}}{S}$ change from 2.08 \% to 1.72 \%. Moreover, the main backgrounds left of this process are channels containing two electrons and two neutrinos accounted for 86\%, which is the same as the final particles of the signal channel.\par
  \begin{table*}[hbtp]
 \caption{Yields for backgrounds and ZH (Z$\rightarrow e^+e^-$, H$\rightarrow$inv.) signal at the CEPC, with $\sqrt{s}$=240 GeV and integrated
luminosity of 5.6 $ab^{-1}$. (Assume BR($H\rightarrow inv.)$=50\%)}
 \label{tab:all channel are grouped  cute}
 \tiny
 \begin{center}
 \renewcommand{\arraystretch}{1.2}
 \begin{tabular}{cccccccccccc}
 \hline
 Process&  eeH\_inv. &2f &single\_w &single\_z &szorsw &zz &ww &zzorww &ZH&total\_bkg&$\frac{\sqrt{S+B}}{S}$ \\ \hline
Total generated & 19712 &801152072 &19517400 &9072951 &1397088 &6389430&50826214 &20440840 &1140495&909936490&153.03 \%   \\
$N_{e^{+}}=1,N_{e^{-}}=1$ & 18405 &389959503 &15669806 &4931933&1236440 &5816250 &47812974 &18467237&679473&484573616& 119.61 \%   \\
120GeV$<M_{recoil}^{e^{+}e^{-}}<$170GeV& 16726 &16124629 &6286116&1272240 &313037 &100041 &9972681 &423152 &35389&34527285& 35.14 \%    \\
71GeV$<M_{e^{+}e^{-}}<$99GeV  & 13677 &5382788 &647494&324692 &113529&15001  &1823446 &92463 &26188 &8425601& 21.24 \%   \\
12GeV$<P_{T}^{e^{+}e^{-}}<$55GeV  & 13134 &3476906&558026 &259411 &98570&12533  &1584475 &79506 &25162&6094589& 18.82 \%    \\
$\Delta\phi^{e^{+}e^{-}}<176$\degree & 12566 &1230398 &516751 &238434&94468 &10493 &1435540 &71282 &24246&3621612& 15.17 \%    \\
103GeV$<E_{visible}<$120GeV & 11618 &4609&30665 &3348 &27463&56  &570 &3430 &131 &70272& 2.46 \%   \\
$1.8<\frac{E_{e^+e^-}}{P_{e^+e^-}}<2.4$& 9654 &1085&14179 &1705 &12209  &10&215 &1127 &61&30591& 2.08 \%     \\
$ReM\_{visdtau}>220$  and Impact\_{Tau}$<$0.0011 & 8641 &442 &1281 &1354 &10244&0 &19 &39 &26  &13405 & 1.72 \% \\
 Efficiency& 43.84 \%& 0.00 \% &0.01 \%&0.01 \% &0.73 \%&0.00 \% &0.00 \%&0.00 \% &0.00 \% &0.00 \%&\\
 \hline
 \end{tabular}
 \end{center}
 \end{table*}
\section{Result of upper limit and the boson mass resolution (BMR)} 	
\begin{figure}[t]
\begin{center}
\centering
\includegraphics[width=0.65\columnwidth]{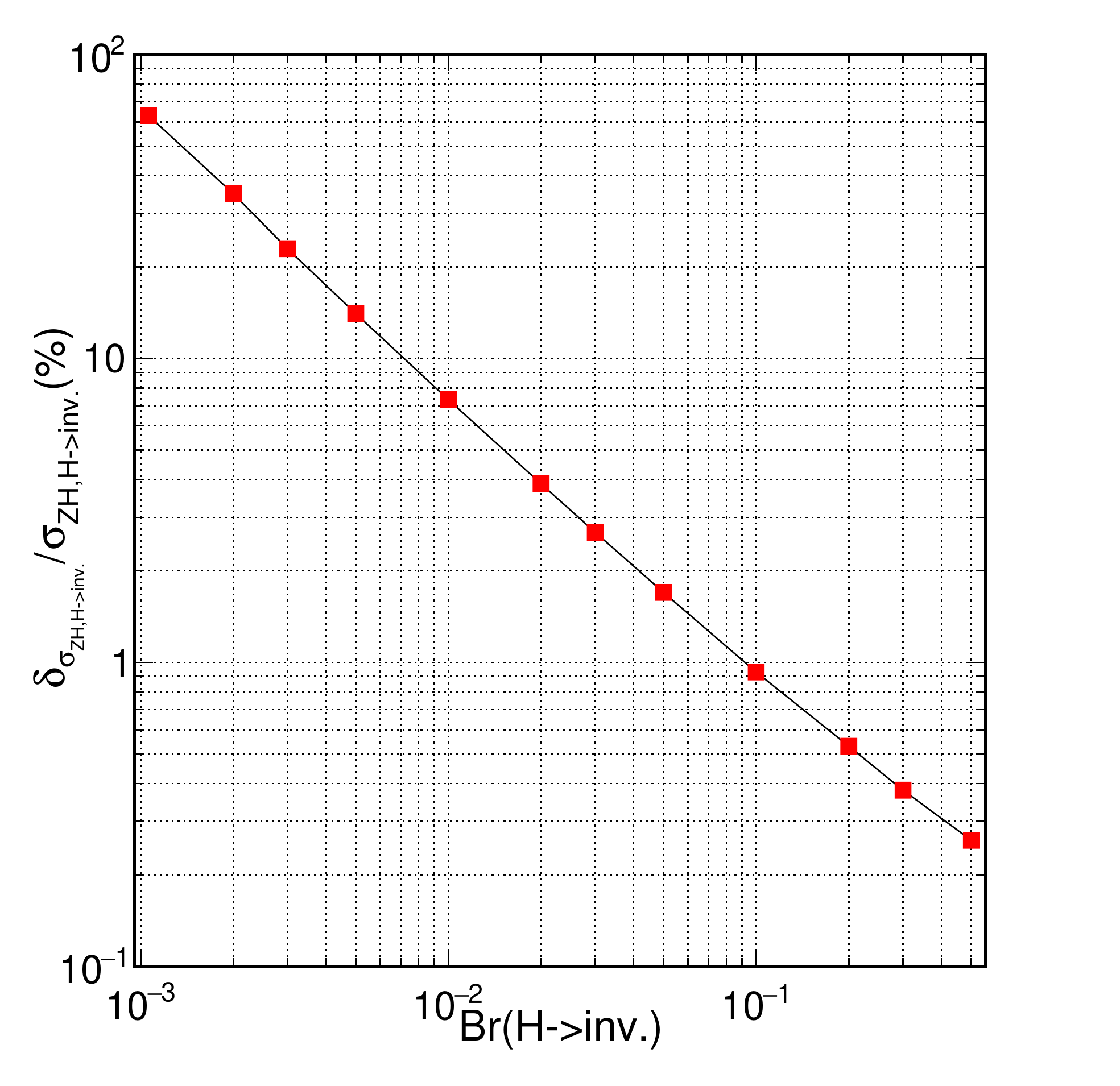}
\caption{The precision of the cross setion of Higgs decay to invisible final states $\delta_{\sigma_{ZH,H\rightarrow inv.}}/\sigma_{ZH,H\rightarrow inv.}$  versus BR(H$\rightarrow$inv.).
\label{fig:higgs_cross_section}}
\end{center}
\end{figure}

\begin{figure}[t]
\begin{center}
\centering
\includegraphics[width=0.6\columnwidth]{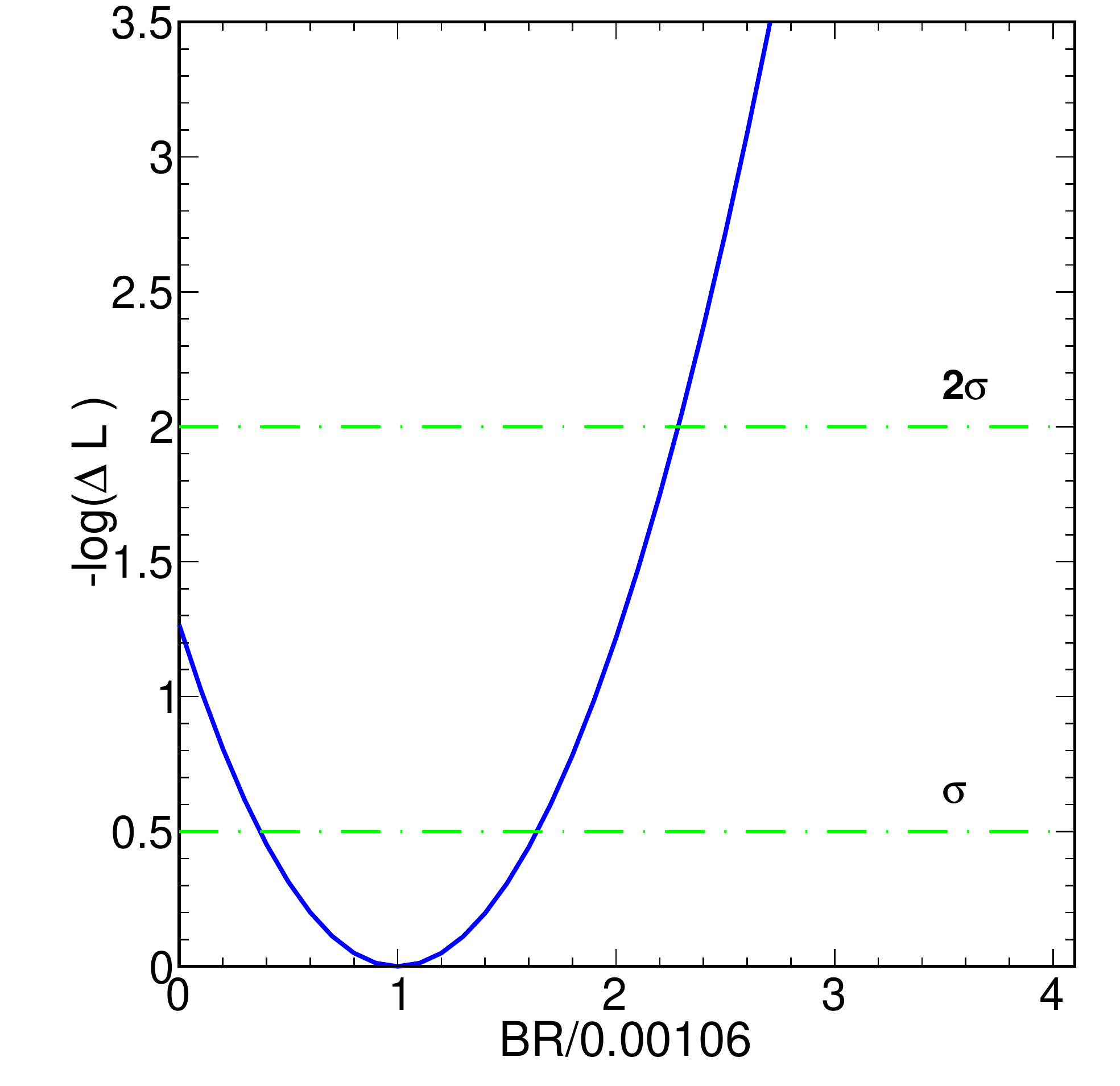}
\caption{(color online) The figure is the likelihood profile of combination, where the green projective line label out the location of 68\%, 95\% confidence level, which corresponds to -$\Delta$ log(L)=0.5, 2 on the y-axis.
\label{fig:likelihood}}
\end{center}
\end{figure}

After the event selections, the 95\%CL upper limit of BR (H$\rightarrow$inv.) is computed using a profile likelihood ratio test statistic~\cite{cowan2011asymptotic} in which systematic uncertainties are ignored. On the event selection, suppose the BR (H$\rightarrow$inv.)=50\% is for the convenience of the signal selection. Based on different assumptions of BR(H$\rightarrow$inv.), the relative precision of $\delta_{\sigma_{ZH,H\rightarrow inv.}}/\sigma_{ZH,H\rightarrow inv.}$ is shown in Fig.~\ref{fig:higgs_cross_section}. The SM value of BR (H$\rightarrow$inv.) is 0.106\%, which will be used to calculate the upper limit of BR (H$\rightarrow$inv.). The likelihood method is $\mu$S+B fitting, where $\mu$ is the signal strength, S is the signal and B is the background. First, we fit the signal and background samples and get the fitting functions. Next, generate Asimov data separately according to the function of signal and background, and the Asimov data can provide a simple method to obtain the median experiment sensitivity of the measurement as well as fluctuations about this expectation. Suppose $\mu$=1 and get the combination Asimov data of $\mu$S+B. Then fit the Asimov data and get the distribution of likelihood profile of $\mu$. The distribution of $\mu$ contains the mean value and the error $\sigma_{\mu}$ of $\mu$.  The combination likelihood profile of three channels is shown in Fig.~\ref{fig:likelihood} where the horizontal axis is $\mu$, and its distribution is the statistical error of the fit. And the horizontal axis corresponding to -$\Delta$ log(L)=2 on the y-axis is the upper limit of BR (H$\rightarrow$inv.) on 95\%C which is estimated to be 0.24\% in Fig.~\ref{fig:likelihood}. \par

Table.~\ref{tab:combin} summarizes the expected precision on the measurement of $\sigma (ZH)\times$BR(H$\rightarrow$inv.) and the 95\% confidence-level (CL) upper limit on BR(H$\rightarrow$inv.) from a CEPC dataset of 5.6$ab^{-1}$. The value of $\sigma (ZH)\times$BR(H$\rightarrow$inv.) is equal to $\sigma_{\mu}$. The combined branching ratio is measured as 0.106\%$\pm$0.067\% and the upper limit at 95\% confidence level is estimated to be 0.24\%. Subtracting the SM $H \rightarrow ZZ^{\ast} \rightarrow \nu\overline{\nu}\nu \overline{\nu}$ contribution, a 95\% CL upper limit of 0.13\% on $BR_{inv.}^{BSM}$ , the BSM contribution to the decay of H$\rightarrow$inv. can be obtained. \par
\begin{figure}[ht]
\begin{center}
\centering
\includegraphics[width=1.0\columnwidth]{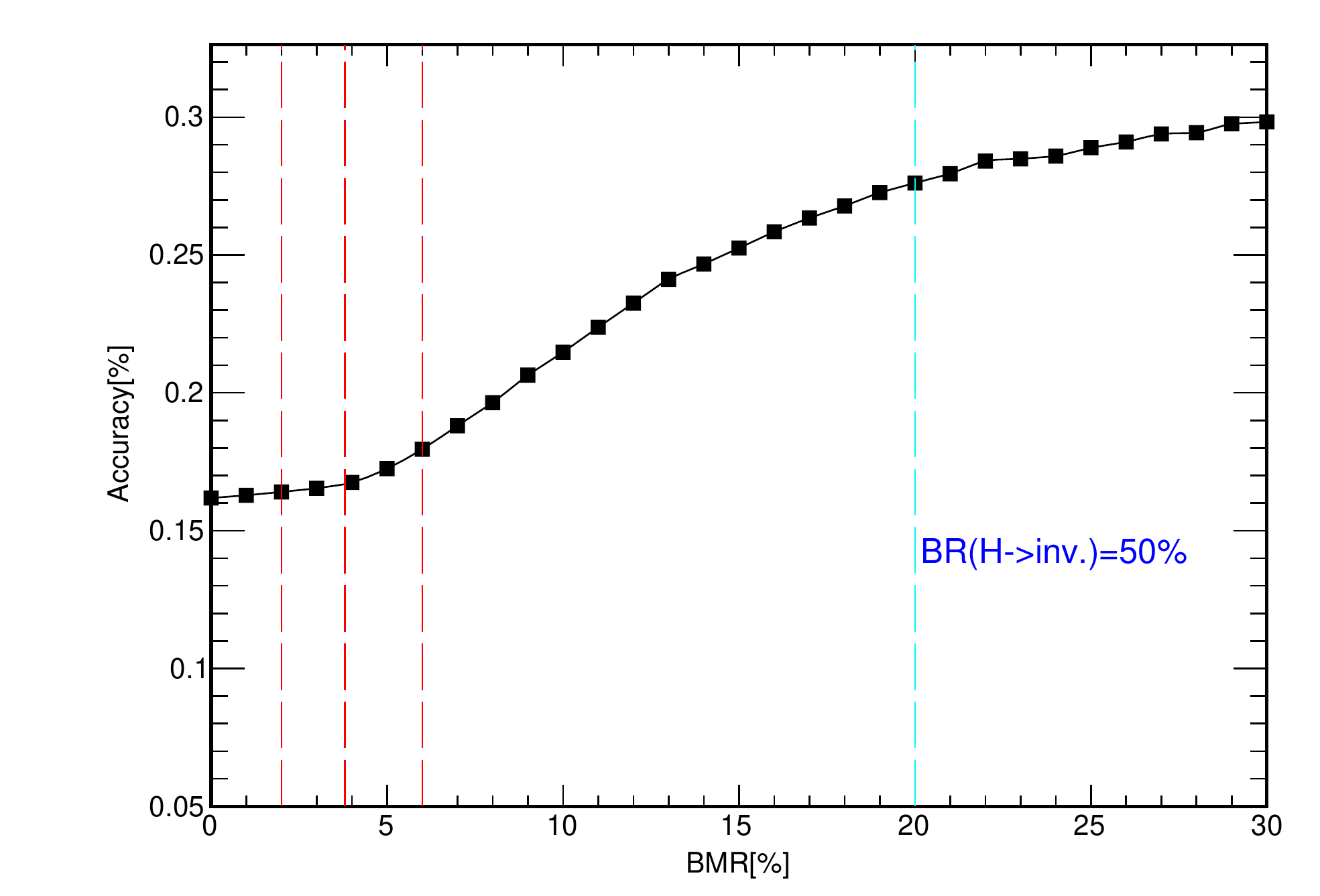}
\caption{(color online) The dash lines show the accuracy when BMR is 2\%, 3.8\%, 6\% and 20\% under assuming the BR (H$\rightarrow$inv.)=50\%. The background is ZZ (Z$\rightarrow$qq, Z$\rightarrow$inv.). This figure indicates the dependence of the accuracy of the qqH (H$\rightarrow$inv.) channel on the boson mass resolution.
\label{fig:BMR}}
\end{center}
\end{figure}

 \begin{table}[htpb]
 \caption{Expected precision on the measurement of $\sigma$(ZH)$\times$BR(H$\rightarrow$inv.) and the 95\% confidence-level (CL) upper limit on BR (H$\rightarrow$inv.) from a CEPC dataset of 5.6$ab^{-1}$.}
 \label{tab:combin}
 \tiny
\begin{tabular}{p{0.15\textwidth}<{\centering}p{0.15\textwidth}<{\centering}p{0.15\textwidth}<{\centering}}%
 \hline
ZH final state studied&  Relative precision on $\sigma(ZH)\times$BR  & Upper limit on BR (H$\rightarrow$inv.) \\
 &&\\ \hline
Z$\rightarrow e^+e^-$, H$\rightarrow$inv.& 403\% & 0.96\%  \\
Z$\rightarrow \mu^+\mu^-$, H$\rightarrow$inv.& 98\% & 0.31\%   \\
Z$\rightarrow q\overline{q}$, H$\rightarrow$inv.& 85\% & 0.29\%   \\ \hline
 Combination& 63\%& 0.24\% \\
 \hline
 \end{tabular}
 \end{table}
The boson mass resolution (BMR) is defined as the resolution of invariant mass of the Higgs at the 240 GeV center-of-mass energy. Base on the CEPC detector, the BMR can reach 3.8\% under Arbor reconstruction algorithm~\cite{Ruan:2014paa,zhao2019higgs}. The final accuracy of the H$\rightarrow$inv. strongly relys on the precision of BMR. Considering the uncertainty of the system, it is necessary to further quantitatively analyze the dependence of BMR. A fast simulation is performed to quantify this dependence. The fast simulation takes into account the qqH (H$\rightarrow$inv.) signal and the background of ZZ (Z$\rightarrow$qq, Z$\rightarrow$inv.), and suppress this background mainly rely on the recoil mass of Higgs. Fig.~\ref{fig:BMR} shows the dependence of the accuracy of qqH (H$\rightarrow$inv.) versus different BMR. For the BMR between 4\% to 20\%, the accuracy degrades rapidly as BMR increases, and for the BMR less than 4\%, the change of accuracy is tiny. For BMR larger than 20\%, the reconstructed invariant and recoil mass of the Z boson can not provide significant separation power, and the degrading tendency saturates. If the BMR degrades from 4\% to 6\%, the Higgs width measurement resolution degrades by 11\%. If the BMR improves to 2\%, the Higgs width measurement resolution only improves 2\%. On CEPC and ILC, BMR is vital for the measurement, which relates to the qqH channel. Therefore, the BMR of less than 4\% will be an essential reference for detector design and optimization.

\section{Conclusion}
This paper investigates the measurement potential of Higgs decay to an invisible channel on CEPC. The upper limit of Higgs invisible decay is measured with a model independent way for three channels, and also get the combined result of a likelihood profile in section IV. Comparing with the LHC results, which is 26\%from ATLAS and 19\% from CMS, the upper limit at 95\% confidence level of the branching ratio of Higgs invisible decay channel on CEPC will be improved as two orders of magnitude. Comparing with other electron-positron colliders  the International Linear Collider (ILC) and the Future Circular Collider (FCC), which the upper limit of BR (H$\rightarrow$inv.) is 0.26\% of ILC~\cite{ishikawa2019search} and 0.22\% of FCC-ee ($5ab^{-1}$ at 240 GeV and 0.19\% by combining 365 GeV)~\cite{deBlas2018}, and the CEPC result is competitive. The statistic error of the CEPC is much lower due to the high luminosity and large statistics. The impact of the BMR on the ZH (Z$\rightarrow$qq,H$\rightarrow$inv.) signal strength accuracy has been studied as well, and the obtained curve shows that the analysis result would be one of the indicators for the CEPC detector optimization study.
\section*{Acknowledgements}
The authors would like to thank Chengdong FU, Gang LI and Xianghu ZHAO for providing the simulation tools and samples. We also thank Patrick Janot for helpful discussion about FCC results. 

\bibliographystyle{elsarticle-num-names}
\bibliography{p4_cepc_higgs_invisible}
\clearpage

\end{document}